\documentclass[
pra,amsmath,amssymb,aps,
twocolumn,
superscriptaddress,
showpacs,
]{revtex4-1}

\usepackage{graphicx}
\usepackage{dcolumn}
\usepackage{bm}

\begin{document}


\title{Predicting and verifying transition strengths from weakly bound molecules}


\author{K. Aikawa}
\email{ka\_cypridina@atomtrap.t.u-tokyo.ac.jp}
\affiliation{Department of Applied Physics, The University of Tokyo, Hongo, Bunkyo-ku, Tokyo 113-8656, Japan}
\author{D. Akamatsu}
\altaffiliation{Present address: National Metrology Institute of Japan, Tsukuba 305-8563, Japan}
\affiliation{Institute of Engineering Innovation, The University of Tokyo, Yayoi, Bunkyo-ku, Tokyo 113-8656, Japan}
\author{M. Hayashi}
\altaffiliation{Present address: NIKON CORPORATION, Kanagawa 252-0328, Japan}
\affiliation{Department of Applied Physics, The University of Tokyo, Hongo, Bunkyo-ku, Tokyo 113-8656, Japan}
\author{J. Kobayashi}
\affiliation{Institute of Engineering Innovation, The University of Tokyo, Yayoi, Bunkyo-ku, Tokyo 113-8656, Japan}
\author{M. Ueda}
\affiliation{JST, ERATO, Yayoi, Bunkyo-ku, Tokyo 113-8656, Japan}
\affiliation{Department of Physics, The University of Tokyo, Hongo, Bunkyo-ku, Tokyo 113-0033, Japan}
\author{S. Inouye}
\affiliation{Institute of Engineering Innovation, The University of Tokyo, Yayoi, Bunkyo-ku, Tokyo 113-8656, Japan}
\affiliation{JST, ERATO, Yayoi, Bunkyo-ku, Tokyo 113-8656, Japan}


\date{\today}

\begin{abstract}
We investigated transition strengths from ultracold weakly bound $^{41}$K$^{87}$Rb molecules 
produced via the photoassociation of laser-cooled atoms. 
An accurate potential energy curve of the excited state $(3)^{1}\Sigma^{+}$ was constructed 
by carrying out direct potential fit analysis of rotational spectra obtained via depletion spectroscopy. 
Vibrational energies and rotational constants extracted from the depletion spectra of $v'=$41--50 levels were 
combined with the results of the previous spectroscopic study, and they were used for modifying 
an \textit{ab initio} potential. An accuracy of 0.14\% in vibrational level spacing and 0.3\% in 
rotational constants was sufficient to predict the large observed variation in transition strengths 
among the vibrational levels. Our results show that transition strengths from weakly bound molecules are a good measure 
of the accuracy of an excited state potential.
\end{abstract}


\maketitle

\section{\label{sec:section1}Introduction}
Ultracold molecular gas is a prominent candidate for realizing novel distinctive applications 
in physics and chemistry, including precision measurements, 
quantum computation, ultracold chemistry, and novel quantum phases \cite{ krems2009cold, carr2009cold}. 
Thus far the production of 
ultracold molecules in the vibrational ground state has been dependent on the optical transfer of 
weakly bound molecules formed via either 
photoassociation \cite{jones2006ultracold} or magnetoassociation \cite{kohler2006production}. 
Among the previously proposed methods 
\cite{sage2005optical, viteau2008optical, deiglmayr2008formation, ni2008high, lang2008ultracold, danzl2010ultracold}, 
stimulated Raman adiabatic passage (STIRAP) \cite{ bergmann1998coherent, vitanov2001coherent, kral2007colloquium} 
of weakly bound molecules is the most efficient method for preparing a molecular sample in a single quantum state. 

\begin{figure}
\includegraphics[width=80mm]{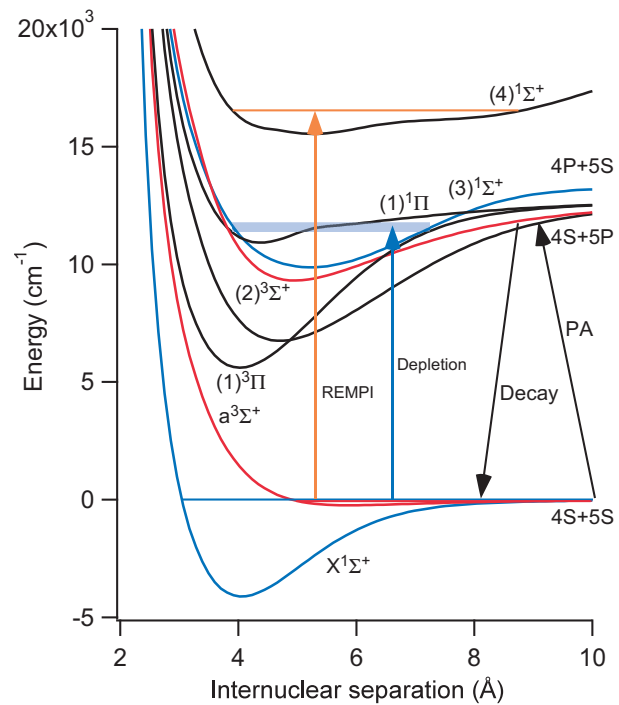}
\caption{\label{fig:KRbPEC}Relevant potential energy curves of KRb. Weakly bound molecules 
were produced by the photoassociation of laser-cooled $^{41}$K and $^{87}$Rb atoms. 
The $(3)^{1}\Sigma^{+}$ state was investigated via the spectroscopy of molecules 
in the $v''=91$ level of $X^{1}\Sigma^{+}$. 
The intermediate state for REMPI ionization was the $(4)^{1}\Sigma^{+}$ state. 
}
\end{figure}

In the STIRAP transfer, an excited state is used as an intermediate state. 
It is important to select an excited state having large 
transition strengths from both weakly bound and deeply bound levels. 
In general, the strengths of molecular transitions from low vibrational levels are readily 
predicted on the basis of the Franck-Condon factors (FCFs) 
calculated from a potential energy curve or molecular constants \cite{herzberg1950molecular}. 
The small number of 
nodes in the radial wavefunction for a low vibrational level indicates a low sensitivity 
to deviations in the radial direction, 
thereby enabling us to explain the intensity distribution over vibrational levels from 
molecular constants. 
However, it is difficult to predict transition strengths from weakly bound molecular 
levels because weakly bound levels have 
a large number of nodes in their wavefunctions; hence, the FCFs are quite sensitive 
to the wavefunctions of excited states. 
In this letter, we show that transition strengths from weakly bound levels can be 
predicted on the basis of an accurate potential energy curve of the excited state, 
constructed via direct potential fit (DPF) analysis \cite{seto2000direct} of both 
vibrational energies and rotational constants. 

We focused on the $(3)^{1}\Sigma^{+}$ state of KRb in the range 11400--11800 
cm$^{-1}$ with respect to the ground atomic threshold, which was proposed as 
a potential candidate for the STIRAP transfer of weakly bound molecules 
to the rovibrational ground state \cite{wang2007rotationally}. An RKR potential 
curve of the $(3)^{1}\Sigma^{+}$ state 
was reported from the potential minimum up to 10400 cm$^{-1}$ \cite{amiot19993, *amiot2000corrigendum}. 
An ionization spectrum for $^{39}$K$^{87}$Rb obtained by using a pulse laser and 
a depletion spectrum near 11700 cm$^{-1}$ is provided in Ref.~\onlinecite{wang2007rotationally}. 
Recently, we have realized the STIRAP transfer of weakly bound molecules ($v''=91$, $J''=0$ of $X^{1}\Sigma^{+}$) 
to the rovibrational ground state ($v''=0$, $J''=0$ of $X^{1}\Sigma^{+}$), 
mediated by the $v'=41$ level of the $(3)^{1}\Sigma^{+}$ state \cite{aikawa2010coherent}. 
Before conducting this experiment, we carried out depletion spectroscopy in the range 11400--11800 cm$^{-1}$, 
which revealed 10 vibrational levels. 
We found that the width of the observed spectra was highly dependent on the vibrational levels of the $(3)^{1}\Sigma^{+}$ state. 
By introducing an analytical representation for power broadening, we extracted the transition strengths for each vibrational level. 
The other electronic states that correlate with the $(3)^{1}\Sigma^{+}$ state via spin-orbit interaction were far away; hence 
we assumed that most of the perturbations from these electronic states were negligible. Thus, we could analyze the experimentally 
obtained spectra on the basis of a single potential curve. 
In addition to the previous spectroscopic work near the bottom of the potential \cite{amiot19993,*amiot2000corrigendum}, 
vibrational energies and rotational constants extracted from the spectra were used to construct an accurate potential 
via DPF analysis. Using the modified potential, it was possible to explain the variation in transition strengths among the vibrational levels 
of the $(3)^{1}\Sigma^{+}$ state in terms of the FCFs. 

\section{Depletion spectroscopy}
Previously, our experimental setup for the spectroscopy of ultracold $^{41}$K$^{87}$Rb molecules was 
described in detail \cite{aikawa2009toward}. We provide a brief summary herein. 
We started with a dual-species magneto-optical trap (MOT) of $1\times10^{8}$ $^{41}$K atoms and $2\times10^{8}$ $^{87}$Rb atoms. 
A compressed MOT (C-MOT) procedure was applied for 40 ms to compress and cool the $^{41}$K and $^{87}$Rb atoms. 
The typical densities and temperatures of $^{41}$K and $^{87}$Rb at the end of C-MOT were $2\times10^{11}$ cm$^{-3}$ and 
400 $\mu$K and $4\times10^{11}$ cm$^{-3}$ and 100 $\mu$K for $^{87}$Rb, respectively. 
A photoassociation (PA) laser (wavenumber, 12570.13 cm$^{-1}$; intensity, $1\times10^{3}$ Wcm$^{-2}$) was applied for 10 ms 
at the end of the C-MOT process. The produced molecules were detected using micro-channel plates (MCP) 
after they were ionized via resonance enhanced multi-photon ionization (REMPI) using a pulsed dye laser (wavenumber, 16543 cm$^{-1}$;
intensity, 3$\times10^{6}$ W cm$^{-2}$). 

Depletion spectra were obtained by monitoring ion counts in the $v''=91$ level of $X^{1}\Sigma^{+}$, 
while a CW Ti:Sapphire laser (Sirah Matisse TX; intensity, 50 Wcm$^{-2}$; beam waist, 350 $\mu$m) was continuously applied and scanned. 
In the present study, we analyzed spectra for the $v''=91$, $J''=2$ level, 
whose binding energy with respect to the atomic threshold $F_{\rm K}$=1+$F_{\rm Rb}$=1 
was measured as $-12.454(1)$ cm$^{-1}$. The frequency of the Ti:Sapphire laser was monitored using a
Fabry-Perot cavity which was locked to $^{87}$Rb D2 line. 
The cavity transmission signal was used to calibrate the variation in the scanning speed and to measure the 
relative frequency of the laser with respect to the cavity transmission peak with a precision of 3 MHz. 
The absolute frequency was measured using a commercial wavemeter (Wavelength WS-7; accuracy, 60 MHz). 
The accuracy of measurements for rotational constants was limited by the spectral width, 
which was of the order of 100 MHz, whereas that for vibrational energies was limited by both the spectral width and the wavemeter.

\begin{figure}
\includegraphics[width=80mm]{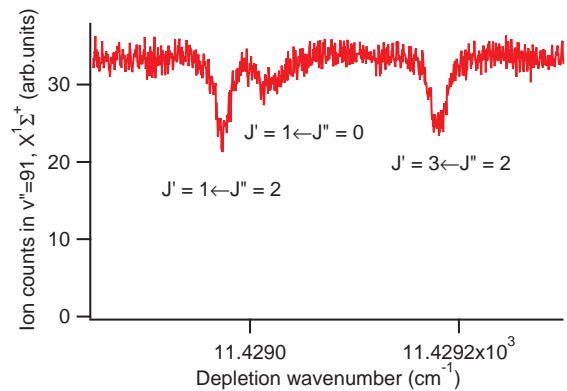}
\caption{\label{fig:Singletdepletion}Depletion spectrum of $v'=41$ level of $(3)^{1}\Sigma^{+}$ taken 
with molecules in $v''=91$ level of $X^{1}\Sigma^{+}$. The spectrum gives rotational constants 
of both ground and excited states. In addition, the width of the spectrum enables us to estimate the transition strength.
}	
\end{figure}

Fig.~\ref{fig:Singletdepletion} shows a depletion spectrum for the $v'=41$ level of the $(3)^{1}\Sigma^{+}$ state, 
which was used as an intermediate state for the STIRAP transfer from 
the $v''=91$, $J''=0$ level of $X^{1}\Sigma^{+}$ to the $v''=0$, $J''=0$ level \cite{aikawa2010coherent}. 
We can extract transition strengths as well as rotational constants for both 
ground and excited states from the spectra. In the appendix, we show that an approximate representation of 
the full-width-half-maximum (FWHM) of a depletion spectrum is given by 
\begin{equation}
2\Delta\sim 0.79\Omega \sqrt{\Gamma \tau}
\label{eq:FWHM}.
\end{equation}
where $\Omega$ is the Rabi frequency; $\Gamma$, the natural width of the excited state; and $\tau$, the duration of spectroscopy. 
Roughly speaking, the width increases not only with the light intensity but also with the duration of spectroscopy. In our case, 
the duration was estimated as a few milliseconds on the basis of the temperature of the molecules and 
beam diameter of the depletion laser. The natural width of the $(3)^{1}\Sigma^{+}$ state was not precisely known, 
but it was obtained as 2$\pi \times 300$ kHz from an ab initio calculation \cite{beuc2006predictions}. 
In the following discussion, we derive the Rabi frequency from the observed spectrum by 
assuming $\tau$ as 2 ms, $\Gamma$ as 2$\pi \times 300$ kHz, and $\sqrt{\Gamma\tau}$ as 60. 

\section{Analysis}
\begin{figure}
\includegraphics[width=80mm]{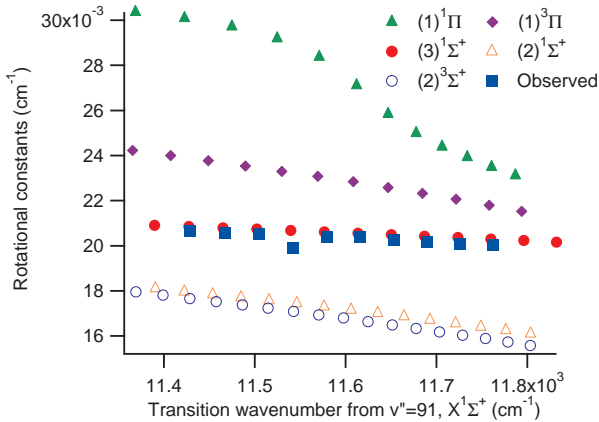}
\caption{\label{fig:Singlet_rotational_constants}Comparison between calculated and observed rotational constants. 
Calculated rotational constants of five molecular states in this range are plotted against energy levels 
from the ground atomic threshold. The observed rotational constants are 
in good agreement with those obtained from the \textit{ab initio} potential of the $(3)^{1}\Sigma^{+}$ state. 
There is a systematic deviation of $\sim$1\% between the observed values and \textit{ab initio} values, 
which indicates the inaccuracy of the \textit{ab initio} potential.
}
\end{figure}
Although the \textit{ab initio} potential \cite{rousseau2000theoretical} enabled us to identify the symmetry of 
the observed spectra without 
any ambiguity, it gave rotational constants that exceeded the  
experimental ones by approximately 1\% (Fig.~\ref{fig:Singlet_rotational_constants}). 
There were two reasons for this deviation. First, the potential minimum exceeded the minimum 
of the RKR curve by 97 cm$^{-1}$. Second, the outer turning point was at the shorter internuclear separation 
than the RKR potential. We found that the variation in the observed transition strengths could not be understood with the \textit{ab initio} potential. 
The observations could not be attributed to the small change in the vibrational quantum number, which corresponded to the energy 
difference in the minimum of the RKR and the \textit{ab initio} potentials. 
These facts indicate that the \textit{ab initio} potential is inaccurate. 
In order to obtain an accurate potential, we carried out DPF analysis, whereby a potential is 
iteratively modified until its eigenvalues coincide with those determined from the experimental spectra \cite{seto2000direct}. 
With the aid of the phiFIT program code \cite{leroy2007phifit}, 
we first constructed an analytical Extended Morse Oscillator (EMO) potential of the 
$(3)^{1}\Sigma^{+}$ state on the basis of (1) the RKR curve, (2) a few points from the inner curve of the \textit{ab initio} 
potential, and (3) a few points around 12500 cm$^{-1}$ from 
the outer curve of the \textit{ab initio} potential. 
The potential curve of the ground state $X^{1}\Sigma^{+}$ was also required for the calculations. 
We used an EMO potential fitted to an accurate, experimentally determined potential \cite{pashov2007coupling}. 
Then, the analytical potential of the $(3)^{1}\Sigma^{+}$ state was modified to reproduce 
our data by using the DPotFit program code \cite{leroy2006dpotfit}. 
A good convergence was achieved when we modified the potential significantly by manually moving the points 
from the \textit{ab initio} potential. This procedure was repeated until the eigenvalues of the potential were within 0.05 cm$^{-1}$ 
of the observed levels, i.e., only 0.14\% of the vibrational level spacing. 
The remaining deviations were presumably due to the incomplete analytical function 
used to represent the potential. Table~\ref{tab:Table_potential} lists the final potential parameters. 
These parameters are used to represent the potential energy curve in the following form:
\begin{eqnarray}
V(R)=&&V_{\rm min}+D_{e}(1-e^{-\phi(R)(R-R_e)})^2\nonumber\\
\phi(R)=&&\sum^{12}_{i=0}\phi_{i}y(R,R_{e})^i\nonumber\\
y(R,R_{e})=&&\frac{R^3-R_{e}^3}{R^3+R_{e}^3}
\label{eq:width}.
\end{eqnarray}

\begin{table}[b]
\caption{\label{tab:Table_potential} Parameters for EMO potential obtained via DPF analysis of the observed spectra. 
The units are cm$^{-1}$ for $V_{min}$ and $D_{e}$, \AA for $R_{e}$, and \AA$^{-1}$ for $\phi_{i}$.}
\begin{ruledtabular}
\begin{tabular}{cccc}
Parameter & Value & Parameter & Value \\ \hline
$V_{\rm min}$	&	9777.6963	&$\phi_5$	&	-0.804673587	\\
$D_{e}$	&	3246.0363	&$\phi_6$	&	3.00474008	\\
$R_{e}$	&	5.25904119	&$\phi_7$	&	11.3877324	\\
$\phi_0$	&	0.449794066	&$\phi_8$	&	-2.140717462	\\
$\phi_1$	&	0.200265883	&$\phi_9$	&	-39.73110586	\\
$\phi_2$	&	0.406840126	&$\phi_{10}$	&	-17.01400311	\\
$\phi_3$	&	0.207884795	&$\phi_{11}$	&	50.93279647	\\
$\phi_4$	&	-0.476349301	&$\phi_{12}$	&	44.83137419	\\
\end{tabular}
\end{ruledtabular}
\end{table}
\begin{figure}
\includegraphics[width=80mm]{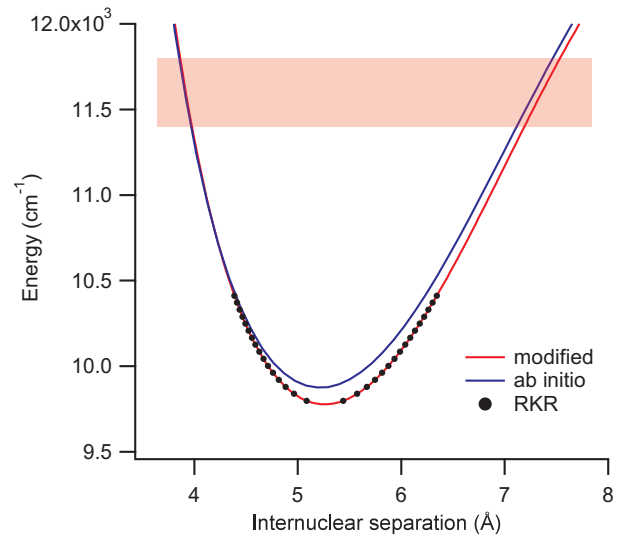}
\caption{\label{fig:Singlet_potential}Potential energy curves of $(3)^{1}\Sigma^{+}$ state. 
The modified potential obtained in the present study is compared with the \textit{ab initio} potential and the experimental RKR potential. 
The shaded area denotes the region where depletion spectra were obtained.
}
\end{figure}
\begin{table*}
\caption{\label{tab:Table_deviations} Comparison between observed, \textit{ab initio}, and modified 
values for vibrational energies (E) and rotational constants (B). 
Vibrational energies are energy levels of the $J'=1$ level. Error bars in vibrational energies are $2\times10^{-3}$ cm$^{-1}$, 
whereas those in rotational constants are $5\times10^{-5}$ cm$^{-1}$. 
Observed vibrational energies are reproduced within $5\times10^{-2}$ cm$^{-1}$, 
whereas observed rotational constants are reproduced within $7\times10^{-5}$ cm$^{-1}$. The vibrational numbering for 
\textit{ab initio} values is deviated by 2 because the potential minimum lies below that of the correct potential.
}
\begin{ruledtabular}
\begin{tabular}{c|ccc|ccc}
$v'$ & \multicolumn{3}{c}{E (cm$^{-1}$)} & \multicolumn{3}{c}{B ($10^{-2}$cm$^{-1}$)} \\ 
      &Obs.&\textit{ab initio} - Obs.&Mod. - Obs.($\times 10^{-3}$)&Obs.&\textit{ab initio} - Obs.($\times 10^{-2}$)&Mod. - Obs. ($\times 10^{-3}$)\\ \hline
41	&	11428.965 	&	-1.5 	&	2.5 	&	2.067	&	1.9	&	-7.0 	\\
42	&	11466.644 	&	-1.6 	&	29.9 	&	2.056	&	2.3	&	-1.7 	\\
43	&	11504.193 	&	-1.8 	&	42.6 	&	2.05	&	2.4	&	-1.5 	\\
44	&	11541.484 	&	-1.9 	&	164.2 	&	1.99	&	7.8	&	52.6 	\\
45	&	11578.919 	&	-2.2 	&	-11.4 	&	2.038	&	2.4	&	-1.5 	\\
46	&	11616.036 	&	-2.4 	&	-28.0 	&	2.037	&	1.8	&	-6.8 	\\
47	&	11652.982 	&	-2.6 	&	-37.3 	&	2.027	&	2.2	&	-3.2 	\\
48	&	11689.751 	&	-2.7 	&	-37.5 	&	2.017	&	2.5	&	0.3 	\\
49	&	11726.308 	&	-2.7 	&	1.2 	&	2.008	&	2.9	&	2.5 	\\
50	&	11762.685 	&	-2.7 	&	43.0 	&	2.005	&	2.5	&	-1.4 	\\
\end{tabular}
\end{ruledtabular}
\end{table*}
Fig.~\ref{fig:Singlet_potential} shows the RKR potential, \textit{ab initio} potential and modified potential. 
At the new potential, rotational constants were reproduced within 0.3\% of the observed values (Table~\ref{tab:Table_potential}). 
We excluded the $v'=44$ level from the analysis because the characteristics of this level are anomalous; 
a much larger transition strength, a much smaller rotational constant, and a much larger deviation in a vibrational energy were observed at this level. 
These features indicate that this level was coupled to the $(2)^{1}\Sigma^{+}$ state 
which was observed at $\sim$1 cm$^{-1}$ above the $v'=44$ level.

\begin{figure}
\includegraphics[width=80mm]{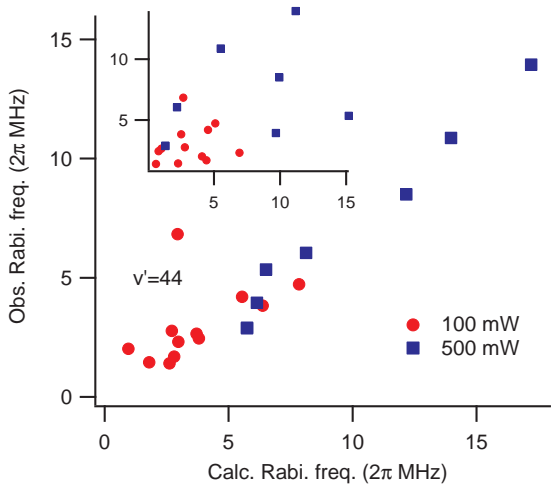}
\caption{\label{fig:Singlet_Rabi} Observed and calculated transition strengths. 
Observed values are determined from the spectra obtained via a depletion laser at 100 mW (red circle) and 500 mW (blue rectangle). 
Calculations based on the modified potential can explain the variation 
in transition strengths with respect to the vibrational levels in terms of the Franck-Condon factors. 
The point deviated from other points corresponds to the $v'=44$ level, which is expected to be mixed with other states.
The inset shows the same plot based on the \textit{ab initio} potential. 
}	
\end{figure}

The most important implication of this analysis is that the variation in the transition strengths 
with respect to the vibrational levels of the excited state can be accurately explained on the basis of the new potential. Fig.~\ref{fig:Singlet_Rabi} 
shows a plot of the Rabi frequencies derived from the observed spectra against those calculated from the corrected potential. 
For comparison, the same plots based on the \textit{ab initio} potential are also shown. 
The calculated values are calibrated on the basis of our recent measurement of the transition dipole moment 
between $(3)^{1}\Sigma^{+}$, $v'=41$ and $X^{1}\Sigma^{+}$, $v''=91$, $0.035(2)ea_{0}$, 
determined via dark resonance spectroscopy of the rovibrational ground-state molecules. 
Weakly bound molecules have more than 90 nodes in their wavefunction; hence, the FCFs are highly dependent on the wavefunction 
of the excited state. In other words, the FCFs can serve as a sensitive measure of the accuracy of wavefunctions. 
Our results show that accuracies of 0.14\% in vibrational level spacings and 0.3\% in 
rotational constants are sufficient to predict the FCFs from weakly bound levels; 
these values are justified by considering the typical size 
of nodes in the radial wavefunction. On one hand, 
the weakly bound level $v''=91$ in the ground state $X^{1}\Sigma^{+}$ has an outer 
turning point of $\sim$10 \AA and an inner turning 
point of $\sim$2 \AA in the internuclear distance. Within these two points, there are 91 nodes; 
hence each node has a typical size of $\sim$0.1 \AA. 
Therefore, the required accuracy for representing the wavefunction is $\sim$10$^{-2}$ \AA. 
On the other hand, in the present analysis, an accuracy of $\sim$0.3\% in rotational constants or $\sim$0.15\% in internuclear 
distance is obtained for the $(3)^{1}\Sigma^{+}$ state because the relation 
between the rotational constant $B$ and the internuclear distance $R$ is given by $B\propto R^{-2}$. 
Assuming the typical size of molecules in the $v'=$41--50 levels of $(3)^{1}\Sigma^{+}$ as 6 \AA, 
we can derive the accuracy of the modified potential in the radial direction as $\sim$10$^{-2}$ \AA; 
this value is in good agreement with the required accuracy for representing the weakly bound level. 

\begin{figure}
\includegraphics[width=80mm]{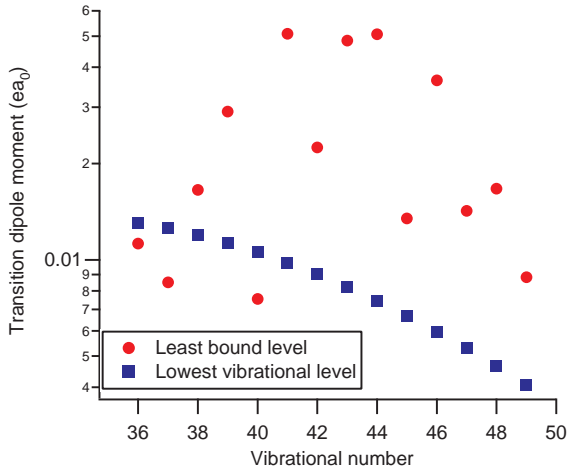}
\caption{\label{fig:singlet_predic} Transition dipole moments of the $(3)^{1}\Sigma^{+}$ state with the $X^{1}\Sigma^{+}$ state 
calculated from the experimentally obtained accurate potential. The absolute values for weakly bound levels are calibrated by 
dark resonance spectroscopy for the transition $v'=41, (3)^{1}\Sigma^{+}\leftarrow v''=91, X^{1}\Sigma^{+}$ whereas 
those for the rovibrational ground state are calibrated by dark resonance spectroscopy for the transition 
$v'=41, (3)^{1}\Sigma^{+}\leftarrow v''=0, X^{1}\Sigma^{+}$.
}	
\end{figure}

Now that we obtained an accurate potential curve as well as the absolute values of the transition dipole 
moment of the $v'=41$, $(3)^{1}\Sigma^{+}$ level with the $v''=0$ and $v''=91$ 
levels of $X^{1}\Sigma^{+}$, we can predict the transition dipole moment for each transition. 
Fig.~\ref{fig:singlet_predic} shows our prediction for the transition dipole moments 
of the $(3)^{1}\Sigma^{+}$ state with the least bound state and for those with 
the lowest rovibrational level $(v''=0)$. The $v'=41$ level used in Ref.~\onlinecite{aikawa2010coherent} 
has favorable wavefunction overlaps with both the weakly bound and the lowest rovibrational levels; 
however other levels such as $v'=38$ and $v'=39$ can potentially serve as an intermediate state 
for the STIRAP transfer of weakly bound molecules to the rovibrational ground state. 
The potential presented herein can enable an accurate prediction for other isotopic combinations of KRb. 
Further, the present method for achieving an accurate potential and verifying its accuracy can be 
extended to other molecular states that exhibit significant spin-orbit mixing by 
evaluating eigenvalues via coupled channel calculations including spin-orbit interaction. 

\section{Conclusion}
The $(3)^{1}\Sigma^{+}$ state of KRb was investigated via the depletion spectroscopy of 
ultracold molecules formed by the photoassociation of laser-cooled $^{41}$K and $^{87}$Rb atoms. 
The spin-orbit mixing of other electronic states 
with the $(3)^{1}\Sigma^{+}$ state was negligible; hence, we could assume this state as a single potential. 
The simplicity of the $(3)^{1}\Sigma^{+}$ state enabled us to modify the potential to reproduce our observations 
as well as to assign the spectra. We observed 10 vibrational levels in the range 11400--11800 cm$^{-1}$ 
with respect to the ground atomic threshold. We developed a theoretical model that related the spectral width with 
the Rabi frequency, which was used to compare the transition strengths for each vibrational level. 
Rotational constants extracted from the observed spectra showed a 1\% deviation from those calculated using 
an \textit{ab initio} potential. By carrying out DPF analysis, we constructed 
an accurate potential that reproduced energy levels with an accuracy of 0.14\% in 
vibrational level spacing and 0.3\% in rotational constants. 
The variation in transition strengths among vibrational levels could be understood in terms of FCFs calculated with the modified potential. 
Our results indicate that the transition strengths from weakly bound levels serve as a sensitive measure 
of wavefunctions, which can be used to test the accuracy of the potential curve. In general,
the proposed procedure can be adopted for constructing an accurate potential and 
verifying its accuracy on the basis of rotational spectra for weakly bound molecules. 

\appendix*
\section{Analytical expression for the line shape of a depletion spectrum}
\begin{figure}
\includegraphics[width=70mm]{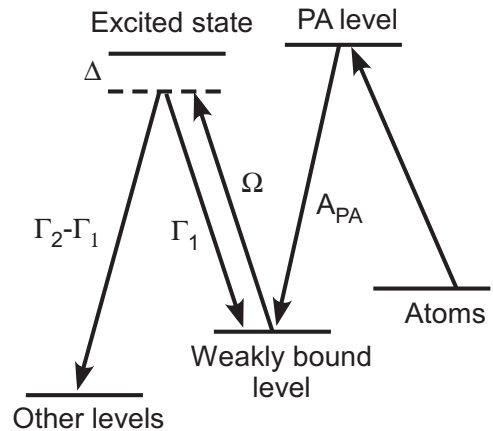}
\caption{\label{fig:depletion} Energy levels relevant to depletion spectroscopy of photoassociated molecules. 
Weakly bound molecules are formed at a rate of $A_{\rm PA}$. A laser having Rabi frequency 
of $\Omega$ detuned from the resonance by $\Delta$ excites ground state molecules. 
Excited molecules spontaneously decay to the initial level and other levels at rates of $\Gamma_1$ 
and $\Gamma_2-\Gamma_1$, respectively. We mainly consider a situation where $\Gamma_{2}>\Gamma_{1}$ is satisfied.
}	
\end{figure}
We consider a general situation, as shown in Fig.~\ref{fig:depletion}, and we assume that radiative transitions occur much faster 
than the time variation of the population in each molecular level because of the photoassociative creation of molecules. 
We first consider the evolution of the population in the ground and excited states of a single molecule. 
The optical Bloch equation for this system can be written as 
\begin{eqnarray}
\frac{ds}{dt}=&&-(\Gamma_2-\Gamma_1)\frac{s-w}{2}\nonumber\\
\frac{dw}{dt}=&&(\Gamma_2+\Gamma_1)\frac{s-w}{2}-2\Omega \mathrm{Im}(\widetilde{\rho_{eg}})\nonumber\\
\frac{\widetilde{\rho_{eg}}}{dt}=&&-(\frac{\Gamma_2}{2}-i\Delta)\widetilde{\rho_{eg}}+\frac{i}{2}w\Omega
\label{eq:OBE}
\end{eqnarray}
where $s=\rho_{gg}+\rho_{ee}$, $w=\rho_{gg}-\rho_{ee}$, and $|e>$ and $|g>$ denote an excited state 
and a weakly bound level, respectively. $\Delta$ and $\Omega$ denote the detuning frequency and the Rabi frequency, respectively. 
The decay rates from the excited state to the initial weakly bound level and to other levels are 
given by $\Gamma_1$ and $\Gamma_2$, respectively. The photoassociation rate is denoted by $A_{PA}$. The width is much larger than the Rabi frequency 
in the experiment; hence, we can assume that the time evolution of $s$ and $w$ is much slower 
than that of $\widetilde{\rho_{eg}}$. Thus, we can set $d\widetilde{\rho_{eg}}/dt=0$ and obtain the following expression for $\widetilde{\rho_{eg}}$:
\begin{equation}
\widetilde{\rho_{eg}}=\frac{i\Omega w}{\Gamma_2-2i\Delta}
\label{eq:rhoeg}
\end{equation}
Substituting Eq.(\ref{eq:rhoeg}) in Eqs.(\ref{eq:OBE}), we obtain alternative equations for $s$ and $w$ as
\begin{eqnarray}
\frac{ds}{dt}=&&-(\Gamma_2-\Gamma_1)\frac{s-w}{2}\nonumber\\
\frac{dw}{dt}=&&(\Gamma_2+\Gamma_1)\frac{s-w}{2}-\frac{2\Omega^{2}\Gamma_{2}w}{\Gamma_2^{2}+4\Delta^{2}} \nonumber\\
\label{eq:OBE2}
\end{eqnarray}
Here, we assume that the time taken by the mean value of the population ratio $w/s$ to attain 
a constant value $z$ after is greater than the typical time for radiative transitions. By using the relation $w=zs$, Eqs.(\ref{eq:OBE2}) give following equations:
\begin{eqnarray}
\frac{ds}{dt}=&&-(\Gamma_2-\Gamma_1)\frac{1-z}{2}s\nonumber\\
z\frac{ds}{dt}=&&(\Gamma_2+\Gamma_1)\frac{1-z}{2}s-\frac{2\Omega^{2}\Gamma_{2}}{\Gamma_2^{2}+4\Delta^{2}}zs \nonumber\\
\label{eq:OBE3}
\end{eqnarray}
Substituting the first equation in the second equation, we obtain a time-independent equation for $z$:
\begin{equation}
(1-z)\left[\Gamma_{2}(1+z)+\Gamma_1(1-z)\right]=\frac{4\Omega^{2}\Gamma_{2}}{\Gamma_{2}^{2}+4\Delta^{2}}z
\label{eq:zexpression}
\end{equation}
This equation is readily solved, and it gives the following expression for $z$.
\begin{eqnarray}
z=&&\frac{\sqrt{(1+k^2)\Gamma_{2}^{2}+2\Gamma_{1}\Gamma_{2}k}-(k\Gamma_2+\Gamma_1)}{\Gamma_2-\Gamma_1}\nonumber\\
k=&&\frac{2\Omega^2}{\Gamma_{2}^{2}+4\Delta^{2}}
\label{eq:OBE4}
\end{eqnarray}
This expression is used in the following discussion. 
Next, we derive rate equations for the population in the ground and excited molecular levels 
for the number of molecules: 
\begin{eqnarray}
\frac{dN}{dt}=&&A_{\rm PA}-(\Gamma_2-\Gamma_1)N_e\nonumber\\
N_e=&&\frac{1-z}{2}N\nonumber\\
N_g=&&\frac{1+z}{2}N
\label{eq:rate}
\end{eqnarray}
where $N_g$ and $N_e$ denote the number of molecules in the ground and excited states, respectively, and $N=N_g+N_e$ is the total number of molecules. 
In these rate equations, a typical timescale is of the order of 1 ms, and it is governed by $A_{\rm PA}$. 
This is much longer than the typical timescale for radiative transitions in most cases ($\leq$1 $\mu$s). The time evolution of $N_g$ is given by
\begin{equation}
\frac{dN_g}{dt}=\frac{1+z}{2}A_{\rm PA}-\frac{1-z}{2}(\Gamma_2-\Gamma_1)N_g
\end{equation}
Thus, the solution for $N_g$ is given by 
\begin{eqnarray}
N_g(\tau)=&&\frac{A_{\rm PA}}{\Gamma_2-\Gamma_1}\frac{1+z}{1-z}\nonumber\\
 &&\times \left(1-\mathrm{exp}\left[-\frac{1-z}{2}(\Gamma_2-\Gamma_1)\tau\right]{}\right)
\label{eq:solution}.
\end{eqnarray}
This expression gives the line shape of a depletion spectrum for a duration $\tau$. Assuming $(\Gamma_2-\Gamma_1)\tau\gg1$, 
Eq.(\ref{eq:solution}) gives 
\begin{equation}
N_g(\tau)\rightarrow\frac{A_{\rm PA}}{\Gamma_2-\Gamma_1}
\end{equation}
on resonance ($\Delta=0$ and $w\rightarrow0$), whereas $N_g$ at an infinite detuning ($\Delta\rightarrow\infty$ and $w\rightarrow1$) 
is given as
\begin{equation}
N_g(\tau)\rightarrow A_{\rm PA}\tau
\end{equation}
$N_g$ at an infinite detuning is much larger than $N_g$ on resonance. Thus, the width of the spectrum is determined by finding $w$ such that it satisfies 
\begin{equation}
N_g(\tau)=\frac{1}{2}A_{\rm PA}\tau
\label{eq:width2}.
\end{equation}
Substituting Eq.(\ref{eq:solution}) in Eq.(\ref{eq:width2}) and rewriting the equation with a new variable $x\equiv1-z(\ll1)$, 
we obtain the following equation:
\begin{equation}
\frac{2}{x}\left(1-\mathrm{exp}\left[-\frac{x}{2}(\Gamma_2-\Gamma_1)\tau\right]\right)=\frac{\Gamma_2-\Gamma_1}{2}\tau
\end{equation}
A rigorous solution of this equation is given by
\begin{eqnarray}
x=&&\frac{4+2{\rm W}(-2/e^2)}{(\Gamma_2-\Gamma_1)\tau}\nonumber\\
=&&\frac{1}{(\Gamma_2-\Gamma_1)\tau}\times 3.18724...
\end{eqnarray}
where W is the Lambert W function. Thus, the parameter $k$ in Eqs.(\ref{eq:OBE4}) is given by
\begin{equation}
k=\frac{3.18724(1-1.59362/\Gamma_2\tau)}{(\Gamma_2-\Gamma_1)\tau-3.18724}
\end{equation}
Assuming $(\Gamma_2-\Gamma_1)\tau\gg1$, we obtain the following expression for FWHM:
\begin{equation}
2\Delta\approx 0.79\Omega\sqrt{(\Gamma_2-\Gamma_1)\tau}
\end{equation}
When decays from the excited state to the initial state are negligible $(\Gamma_2\gg\Gamma_1)$, we obtain a simple relation (\ref{eq:FWHM}). 
It is difficult to evaluate the numerical factor $\sqrt{\Gamma\tau}$ precisely; 
therefore, the width of a depletion spectrum cannot serve as an accurate measure of the transition strength. 
However, the expression (\ref{eq:FWHM}) enables us to systematically compare transition strengths for different vibrational levels.

\begin{acknowledgments}
We thank P. Naidon and T. Kishimoto for insightful discussions, and K. Oasa, Y. Tanooka, and K. Mori 
for their assistance with the experiment. K. A. and D. A. acknowledge the support of the Japan Society for the Promotion of Science.
\end{acknowledgments}

\providecommand{\noopsort}[1]{}\providecommand{\singleletter}[1]{#1}%

\end{document}